\documentclass[english,reprint, superscriptaddress, breaklinks=true, showkeys, showpacs=false, nofootinbib]{revtex4}

\usepackage[T1]{fontenc}
\usepackage{color}
\usepackage{babel}
\usepackage{tikz}
\usepackage{float}
\usepackage{amssymb}
\usepackage{listings}
\usepackage{graphicx}
\usepackage{amsmath}
\usepackage{svg}

\usepackage{tikz}
\usetikzlibrary{quantikz}
\usepackage{algorithm2e}

\usepackage{algpseudocode}

\usepackage[unicode=true,pdfusetitle, bookmarks=true,bookmarksnumbered=false,bookmarksopen=false, breaklinks=true,pdfborder={0 0 0},backref=false,colorlinks=true]{hyperref}
\usepackage{breakurl}
\definecolor{Code}{rgb}{0,0,0}
\definecolor{Decorators}{rgb}{0.5,0.5,0.5}
\definecolor{Numbers}{rgb}{0.5,0,0}
\definecolor{MatchingBrackets}{rgb}{0.25,0.5,0.5}
\definecolor{Keywords}{rgb}{0,0,1}
\definecolor{self}{rgb}{0,0,0}
\definecolor{Strings}{rgb}{0,0.63,0}
\definecolor{Comments}{rgb}{0,0.63,1}
\definecolor{Backquotes}{rgb}{0,0,0}
\definecolor{Classname}{rgb}{0,0,0}
\definecolor{FunctionName}{rgb}{0,0,0}
\definecolor{Operators}{rgb}{0,0,0}
\definecolor{Background}{rgb}{0.98,0.98,0.98}
\lstdefinelanguage{Python}{
numbers=left,
numberstyle=\footnotesize,
numbersep=1em,
xleftmargin=1em,
framextopmargin=2em,
framexbottommargin=2em,
showspaces=false,
showtabs=false,
showstringspaces=false,
frame=l,
tabsize=4,
basicstyle=\ttfamily\small\setstretch{1},
backgroundcolor=\color{Background},
commentstyle=\color{Comments}\slshape,
stringstyle=\color{Strings},
morecomment=[s][\color{Strings}]{"""}{"""},
morecomment=[s][\color{Strings}]{'''}{'''},
morekeywords={import,from,class,def,for,while,if,is,in,elif,else,not,and,or,print,break,continue,return,True,False,None,access,as,,del,except,exec,finally,global,import,lambda,pass,print,raise,try,assert},
keywordstyle={\color{Keywords}\bfseries},
morekeywords={[2]@invariant,pylab,numpy,np,scipy},
keywordstyle={[2]\color{Decorators}\slshape},
emph={self},
emphstyle={\color{self}\slshape},
}

\makeatletter
\@ifundefined{textcolor}{}
{%
 \definecolor{BLACK}{gray}{0}
 \definecolor{WHITE}{gray}{1}
 \definecolor{RED}{rgb}{1,0,0}
 \definecolor{GREEN}{rgb}{0,1,0}
 \definecolor{BLUE}{rgb}{0,0,1}
 \definecolor{CYAN}{cmyk}{1,0,0,0}
 \definecolor{MAGENTA}{cmyk}{0,1,0,0}
 \definecolor{YELLOW}{cmyk}{0,0,1,0}
}

\hypersetup{colorlinks=true,citecolor=blue,linkcolor=cyan,urlcolor=blue,filecolor= green, breaklinks=true}
\usepackage{url}
\usepackage{breakurl}

\makeatother

\begin{document}

\SetKwComment{Comment}{/* }{ */}

\title{Evolution strategies: Application in hybrid quantum-classical neural networks}

\author{Lucas Friedrich}
\email[Electronic address: ]{lucas.friedrich@acad.ufsm.br}
\affiliation{Physics Departament, Center for Natural and Exact Sciences, Federal University of Santa Maria, Roraima Avenue 1000, 97105-900, Santa Maria, RS, Brazil}

\author{Jonas Maziero}
\email[Electronic address (corresponding author): ]{jonas.maziero@ufsm.br}
\affiliation{Physics Departament, Center for Natural and Exact Sciences, Federal University of Santa Maria, Roraima Avenue 1000, 97105-900, Santa Maria, RS, Brazil}


\begin{abstract}

With the rapid development of quantum computers, several applications are being proposed for them. Quantum simulations, simulation of chemical reactions, solution of optimization problems and quantum neural networks (QNNs) are some examples. However, problems such as noise, limited number of qubits and circuit depth, and gradient vanishing must be resolved before we can use them to their full potential. In the field of quantum machine learning, several models have been proposed. In general, in order to train these different models, we use the gradient of a cost function with respect to the model parameters. In order to obtain this gradient, we must compute the derivative of this function with respect to the model parameters. One of the most used methods in the literature to perform this task is the parameter-shift rule method. This method consists of evaluating the cost function twice for each parameter of the QNN. A problem with this method is that the number of evaluations grows linearly with the number of parameters. In this work we study an alternative method, called  Evolution Strategies (ES), which are a family of black box optimization algorithms which iteratively update the parameters using a search gradient. An advantage of the ES method is that in using it one can control the number of times the cost function will be evaluated. We  apply the ES method to the binary classification task, showing that this method is a viable alternative for training QNNs. However, we observe that its performance will be strongly dependent on the hyperparameters used. Furthermore, we also observe that this method, alike the parameter shift rule method, suffers from the problem of gradient vanishing.

\end{abstract}


\maketitle

\section{Introduction}

Many developments in science and technology in the last years were obtained with the aid of artificial intelligence. Its applications extend to the most varied areas of knowledge, such as for example computer vision \cite{computer_vision_1,computer_vision_2,computer_vision_3}, natural language processing \cite{nlp_1,nlp_2}, drug discovery \cite{drog_discovery}, analysis of astronomical images \cite{dl_astronomia}, and chemistry simulations \cite{Deep_Chemistry}. With the development of quantum computers, several studies are being conducted with the aim of taking artificial intelligence to the quantum domain \cite{7,7.2,7.3,7.4,7.5,7.6,7.7,7.8}. It is hoped that, by utilizing phenomena such as entanglement and superposition, we will be able to create models more powerful than their classical counterparts.

Models such as Quantum Multilayer Perceptron \cite{quantum_model_multilayer_perception}, Quantum Convolutional Neural Networks \cite{qunatum_Convolutional}, Quantum Kernel Method \cite{kernel_methods}, and Quantum-Classical Hybrid Neural Networks (HQCNN) \cite{hybrid_1,hybrid_2,hybrid_3,hybrid_4} are some candidate models. In the era of noisy intermediate-scale quantum devices (NISQ), hybrid models are the most used. This era is characterized by the limited number of qubits we have access to and the presence of noise. Hybrid models are built using sequential classical and quantum layers. With this, we are able to create models with fewer qubits. 
For training these models, the gradient descent method or its variants has been used in the literature. The gradient descent method consists of using the gradient of a cost function to update the parameters of the neural network. To obtain the gradient of the quantum layers, one can use the method called parameter-shift rule (PSR), which consists of evaluating the cost function for each parameter of the quantum layers \cite{maria,11}. In the NISQ era, where the number of qubits that we have access to is limited and the depth of the parameterizations is also limited by noise and decoherence, the use of the PSR method is the most indicated because with it we are able to obtain the derivatives of the cost function analytically. However, as quantum computers develop, increasing depth and qubit numbers make this method impractical. Therefore, it is necessary to
find alternative ways for the task of training.

Evolution strategies (ES) \cite{es_1, es_2,10} are a family of black box optimization algorithms. These algorithms have already been applied to a variety of classical problems. For example, ES has been shown to be an alternative to reinforcement learning \cite{12.3}. However, application in the quantum domain is still understudied, with only a few works  \cite{12.0, 12.1, 12.2} using these methods. Such a strategy is promising, because the number of evaluations of the cost function does not scale with the number of parameters. Furthermore, as the cost function evaluations are independent, this method can be parallelized. However, its parallelized application in quantum computing is still limited in the NISQ era. Finally, we should note that this method will also be strongly influenced by the number of times the cost function will be evaluated, that is, the value of $\lambda$ used, because the lower this value, the lower the accuracy of this method.

This article is organized as follows. In Sec. \ref{hqcnn}, we review the HQCNN models, where a parallel is made with the classical deep neural network models. After that, in Sec. \ref{Classical Layer}, we briefly discuss how classical layers work. In Sec. \ref{Quantum Layer}, we describe how a quantum layer works, and in Sec. \ref{Encoder} we discuss different ways of mapping classical data into quantum states. In the sequence, in Sec. \ref{Parametrization}, we discuss how the parameterization of a quantum layer is done. Finally, in Sec. \ref{Measurements},
we describe how measurements can be made on HQCNN. In Sec. \ref{Natural Evolutionary Strategies}, we present a review of the ES method, showing how to estimate the gradient of a given function. In Sec. \ref{Method}, we start by describing the two HQCNN models that we use in this study. In Sec. \ref{Treinamento}, we show how a HQCNN model is trained, and we present our training proposal using ES. In Sec. \ref{Experiments}, we briefly describe how the simulations are done, and, in Sec. \ref{Results}, we present the results we obtained. Finally, in Sec. \ref{Conclusions},
we give our conclusions.

\section{Quantum-Classical Hybrid Neural Networks}
\label{hqcnn}

Classical deep neural network models are in great extent responsible for the success of artificial intellegence in the last few decades. These networks are characterized by the use of several concatenated classical layers. That is to say, for a model with depth $d$, we have
\begin{equation}
    C = L_{ n_{d-1} \rightarrow  n_{d} } \circ L_{ n_{d-2} \rightarrow  n_{d-1} } \circ \cdot \cdot \cdot \circ  L_{ n_{1} \rightarrow  n_{2} } \circ  L_{ n_{0} \rightarrow  n_{1} },
\end{equation}
where each $L$ represents a classical layer and the first and second indices represent the input size and the output size respectively. HQCNN models are also characterized by the use of several concatenated layers, that is
\begin{equation}
    \mathcal{Q} = \mathcal{L}_{ n_{d-1} \rightarrow  n_{d} } \circ \mathcal{L}_{ n_{d-2} \rightarrow  n_{d-1} }\circ \cdot \cdot \cdot \mathcal{L}_{ n_{1} \rightarrow  n_{2} } \circ \mathcal{L}_{ n_{0} \rightarrow  n_{1} }, \label{eq:10000}
\end{equation}
where each $\mathcal{L}$ represents a classical or quantum layer.
We can see that the difference between classical deep neural network models and hybrid quantum-classical models is due to the addition of quantum layers to the classical model. It is expected that by using quantum layers together with classical layers we will be able to build models with greater power and accuracy than models using only classical layers.

\subsection{Classical Layer} \label{Classical Layer}

The structures known as \emph{layers} are one of the main building blocks of all modern classical deep neural network models. Such structures map an input of dimension $n_{0}$ to an output of dimension $n_{1}$. A typical example of these structures is a linear transformation followed by a nonlinear activation function, defined by
\begin{equation}
     \mathcal{L}_{ n_{0} \rightarrow  n_{1} } = \phi( \pmb{W}\pmb{x}+\pmb{b} ), \label{eq:100.1}
\end{equation}
where $\pmb{x}$ is an input vector with dimension $n_{0}$, $\pmb{W}$ is a matrix with dimensions $ n_{1} \times n_{0} $ and $\pmb {b}$ is a vector with dimension $n_{1}$. The elements of $\pmb{W}$ are real values that are updated throughout training. One of the key pieces of deep neural network models is the nonlinearity implemented by the $\phi$ function. There are several functions that we can use to apply nonlinearity, such as the hyperbolic tangent, the Sigmoid or the ReLu.

Furthermore, we can mention the neural network architecture with convolutional layers, which are mainly applied to problems involving computer vision, and the Long short-term memory (LSTM), which are used when dealing with problems related to time series, among many others.

\subsection{Quantum Layer} 
\label{Quantum Layer}

\subsubsection{Encoder} 
\label{Encoder}

The first task when building a quantum layer is to encode the data of interest in quantum states. For this, we can use different strategies, such as the wave function encoder
\begin{equation}
   | \pmb{x} \rangle := \frac{1}{ \Vert \pmb{x} \Vert_{2}^{2} }\sum_{i=1}^{2^{N}}x_{i}|i \rangle, \label{eq:14}
\end{equation}
the dense angle coding
\begin{equation}
   | \pmb{x} \rangle = \bigotimes_{i=1}^{N/2}\cos( \pi x_{2i-1} )|0\rangle  + e^{2\pi i x_{2i}}\sin( \pi x_{2i-1} )|1\rangle,  \label{eq:15}
\end{equation}
or the qubit encoding
\begin{equation}
    |\pmb{x}\rangle = \bigotimes_{i=1}^{N} \cos(x_{i})|0\rangle + \sin(x_{i})|1\rangle.
    \label{eq:16}
\end{equation}
The performance of quantum layers is influenced by this choice \cite{dataEncoder_1,dataEncoder_2,dataEncoder_4}. For example it was shown, in Ref. \cite{dataEncoder_2}, that if we reload the data between the different layers of the parameterization, we will be able to create a model with greater classification capability.

For the purposes of this work we use two alternative forms of data encoding, Figs. \ref{fig:3} and \ref{fig:4}. Furthermore, in subsections model 1 and model 2, Figs. \ref{fig:1} and \ref{fig:122} show which parameterization is used for each case.

\subsubsection{Parameterization}
\label{Parametrization}

After mapping the data of interest to a quantum state, the next step is to apply a parameterization. We write our parameterization as
\begin{equation}
   U(\pmb{\theta})= \prod_{i=1}^{L}U_{i}W_{i}, \label{eq:17}
\end{equation}
with 
\begin{equation}
   U_{i}= \bigotimes_{j=1}^{N} R_{\sigma}^{j,i}(\theta_{j,i}),
\end{equation}
where  $ R_{\sigma}^{i,j}(\theta_{j,i}) = e^{-i \theta_{j,i} \sigma/2} $ with $\sigma \in \{ \sigma_{x},\sigma_{y},\sigma_{z}, \} $ being one of the Pauli matrices and $W_{i}$ are unparameterized gates. 

There are several ways to build the parameterization, and different parameterizations have different expressiveness. Expressiveness is defined as the ability of a given parameterization to explore the Hilbert space. However, the relationship between this choice and the performance of the model is not direct. 

\subsubsection{Measurements}
\label{Measurements}

After mapping the data of interest to a quantum state and applying the parameterization $U(\pmb{\theta})$, Eq. \eqref{eq:17}, the next step is to perform the measurements. The measurements can be done globally, where all qubits are measured, or locally, where only a few qubits are measured individually or in pairs. As a result of these measurements, we can estimate the mean value
\begin{equation}
    f_{i} = Tr[ A_{i} \rho_{x}^{out} ], \label{eq:17.1}
\end{equation}
where $A_{i}$ is a Hermitian operator and $ \rho_{x}^{out} $ is the density matrix at the end of the quantum circuit. That is, given an input $| \pmb{x} \rangle $ and a parameterization $U(\pmb{\theta})$, Eq. \eqref{eq:17}, we have that $ \rho_{x}^{out} = U(\pmb{\theta})|\pmb{x} \rangle \langle \pmb{x}|U(\pmb {\theta})^{\dagger} $.

A particular case is
\begin{equation}
    A_{i} = |i \rangle\langle i|.
\end{equation}
In this case, the outputs will be the respective probabilities of our circuit being in a state of the computational basis. This is a definition of what we can call a global measurement, where all qubits contribute to the value of $f_{i}$.

Another case is when we define
\begin{equation}
    A_{i} = \mathbb{I}_{\bar{i}} \otimes |0\rangle \langle0|_{i}.
    \label{eq:17.2} 
\end{equation}
Here the index $\bar{i}$ indicates that the identity operator will be applied to all qubits with exception to the qubit with index $i$. The index $i$ indicates that $|0\rangle \langle0|$ will be applied only to the qubit of index $i$. This is a definition for what we call local measurements, where only the qubit with index $i$ contributes to the value of $f_{i}$. In this specific case, where $|0\rangle \langle0|$ is used, we will get the probability for the qubit $i$ to be in the state $|0\rangle$.

\section{Evolution Strategies}
\label{Natural Evolutionary Strategies}

Given a function $ f(\pmb{z}), $ with $ \pmb{z} \in \mathbb{R}^{d} $, in the evolution strategy (ES) we reparameterize this function as follows:
\begin{equation}
    J(\theta) = \mathbb{E}_{\theta}[f(\pmb{z})] = \int f(\pmb{z}) \pi(\pmb{z}|\theta)d\pmb{z}. \label{eq:1}
\end{equation}
By differentiating Eq. \eqref{eq:1} with respect to $ \theta $, we get
\begin{equation}
\begin{split}
    \nabla_{\theta}J(\theta) & = \nabla_{\theta} \int f(\pmb{z})\pi(\pmb{z}|\theta)d\pmb{z} \\
    &=\int[ f(\pmb{z}) \nabla_{\theta}\log{\pi(\pmb{z}|\theta)} ]\pi(\pmb{z}|\theta)d\pmb{z}\\
    &=\mathbb{E}_{\theta}[f(\pmb{z})\nabla_{\theta}\log{\pi(\pmb{z}|\theta)} ].
    \label{eq:2}
\end{split}
\end{equation}
For more details about the derivation given in Eq. \eqref{eq:2}, see Ref. \cite{10}.

Thus, we can estimate the gradient using
\begin{equation}
    \nabla_{\theta}J(\theta) \approx \frac{1}{\lambda}\sum_{k=1}^{\lambda} f(\pmb{z}_{k})\nabla_{\theta}\log{ \pi(\pmb{z}_{k}|\theta) },
    \label{eq:3}
\end{equation}
where $\lambda$ is the number of random samples in the parameter space. 
For the case of a Gaussian distribution we have
\begin{equation}
    \pi( \pmb{z} | \theta ) = \frac{1}{ \sqrt{(2\pi)^{d} \det{\pmb{\Sigma}}} }\exp{\Big( -\frac{1}{2}(\pmb{z}-\pmb{\mu})^{T}\pmb{\Sigma}^{-1}(\pmb{z}-\pmb{\mu}) \Big)},
    \label{eq:4}
\end{equation}
where $\pmb{\mu}$ is a mean vector and $\pmb{\Sigma}$ is the covariance matrix. So the parameter $\theta$ is defined as $ \theta := \{ \pmb{\mu}, \pmb{\Sigma} \} $. Therefore, we have that $ \nabla_{\theta}\log{ \pi(\pmb{z}|\theta) } $ will be given by
\begin{equation}
    \nabla_{\pmb{\mu}}\log{ \pi(\pmb{z}|\theta) } = \pmb{\Sigma}^{-1}(\pmb{z}-\pmb{\mu})
    \label{eq:5}
\end{equation}
and
\begin{equation}
    \nabla_{\pmb{\Sigma}}\log{ \pi(\pmb{z}|\theta) } = \frac{1}{2}\pmb{\Sigma}^{-1}(\pmb{z}-\pmb{\mu})(\pmb{z}-\pmb{\mu})^{T}\pmb{\Sigma}^{-1}-\frac{1}{2}\pmb{\Sigma}^{-1}.  \label{eq:6}
\end{equation}

The main ES methods use both Eq. \eqref{eq:5} and Eq. \eqref{eq:6} to obtain an estimate of Eq. \eqref{eq:3}. In this study we will use $\pmb{\Sigma} = \sigma^{2}I$, with $\sigma$ being a constant.  So we have
\begin{equation}
    \nabla_{\theta}J(\theta) \approx \frac{1}{\lambda \sigma^{2}}\sum_{k=1}^{\lambda}(\pmb{z}_{k}-\mu) \cdot f(\pmb{z}_{k}), \label{eq:7}
\end{equation}
with $ \pmb{z}_{k} \sim\mathcal{N}(\pmb{\mu},\sigma^{2}I)$. The algorithm that implements Eq. \eqref{eq:7} is shown below.

\begin{algorithm}[H]
    \caption{Gradient estimation using Eq.  \eqref{eq:7}. }
    \label{alg:1}
    \SetKwInput{KwInput}{Input}                
    \SetKwInput{KwOutput}{Output}  
     \KwInput{ $f(z)$, $\theta_{i}$, $\sigma$ }
     \KwOutput{ $\nabla_{\theta}J(\theta)$ }

     \For{ $k=1...\lambda$ }{
     
        $  z_{k} \sim  \mathcal{N}(\mu,\sigma^{2}I)  $ 
        $ f(z_{k}) $ \Comment*[r]{Evaluates function $f$ in $z_{k}$} 
        $ \nabla_{\theta}J(\theta) = (z_{k}-\mu) $ \Comment*[r]{Calculates the derivative} 
     }
     $ \nabla_{\theta}J(\theta)  = \frac{1}{\lambda\sigma^{2}}\sum_{k=1}^{\lambda}(z_{k}-\mu) \cdot  f(z_{k}) $ \Comment*[r]{Estimates the gradient}  
     
    
\end{algorithm}

\section{Method} 
\label{Method}

\subsection{Models}

For this study, we will use two HQCNN models. The first model will be built using two quantum layers. The second model will be built using two classical layers and a quantum layer. In this work, we define the cost function as
\begin{equation}
    C(\Delta) := \frac{1}{N} \sum_{j=1}^{N}(y_{i}^{j}-\bar{y}_{i}^{j})^{2},
    \label{eq:200}
\end{equation}
where $y_{i}$ is the vector obtained at the end of the network given the input $x_{i}$ and $\bar{y}_{i}$ is the respective desired output. Here $\Delta :=\{\pmb{W},\pmb{\theta}\} $ is defined as the set of network parameters.

\subsubsection{Model 1} 
\label{Modelo 1}

\begin{figure}[H]
    \centering
    \begin{quantikz}
    \qw&\gate[3][0.5cm]{INPUT_{1}}&\qw\\
    \qw&&\qw\\
    \qw&&\qw\\
    \end{quantikz}
     = \begin{quantikz}[row sep=0.2cm,column sep=0.1cm]
    \qw&\qw&    \gate[1]{ R_{y}(x_{1}) }  & \ctrl{1} & \qw     &  \qw      &  \qw       & \gate[1]{ R_{y}(x_{4}) }  & \ctrl{1} & \qw     &  \qw      &  \qw       & \gate[1]{ R_{y}(x_{7}) }\\
    \qw&\qw&    \gate[1]{ R_{y}(x_{2}) }  & \targ{}  & \ctrl{1}&  \qw      &   \qw      & \gate[1]{ R_{y}(x_{5}) } & \targ{}  & \ctrl{1}&  \qw      &   \qw      & \gate[1]{ R_{y}(x_{8}) } \\
    \qw&\qw&    \gate[1]{ R_{y}(x_{3}) }  & \qw      & \targ{} &  \qw &    \qw     & \gate[1]{ R_{y}(x_{6}) } & \qw      & \targ{} &  \qw &    \qw     & \gate[1]{ R_{y}(x_{9}) } \\
    \end{quantikz}
    \caption{ $INPUT_{1}$ for the case of three qubits.}
    \label{fig:3}
\end{figure}
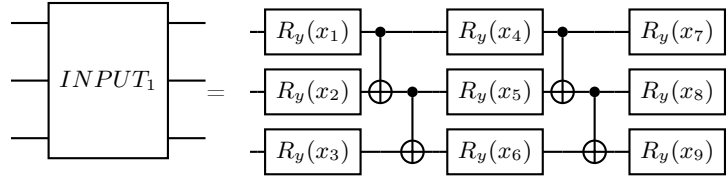

The first layer will be built using five qubits, as illustrated in Fig. \ref{fig:1}. Data will be mapped using real amplitudes as illustrated in Fig. \ref{fig:3}. This is a parameterization used for machine learning and quantum chemistry problems. The parameterization will be given by $U(\pmb{\theta})$, Eq. \eqref{eq:17}. In Fig. \ref{fig:2} shows how the parameterization is done for each $U_{i}$. For measurements we will use Eq. \eqref{eq:17.1} with the Hermitian operator defined in Eq. \eqref{eq:17.2} for the last three qubits.

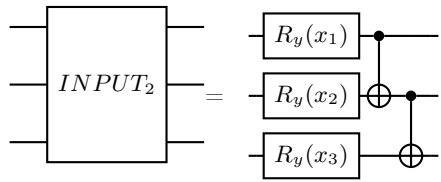
\begin{figure}[H]
    \centering
    \begin{quantikz}
    \qw&\gate[3][0.5cm]{INPUT_{2}}&\qw\\
    \qw&&\qw\\
    \qw&&\qw\\
    \end{quantikz}
     = \begin{quantikz}[row sep=0.2cm,column sep=0.1cm]
    \qw&\qw&    \gate[1]{ R_{y}(x_{1}) }  & \ctrl{1} & \qw     &  \qw      &  \qw       \\
    \qw&\qw&    \gate[1]{ R_{y}(x_{2}) }  & \targ{}  & \ctrl{1}&  \qw      &   \qw       \\
    \qw&\qw&    \gate[1]{ R_{y}(x_{3}) }  & \qw      & \targ{} &  \qw &    \qw     \\
    \end{quantikz}
    
    \caption{ $INPUT_{2}$ for the case of three qubits.}
    \label{fig:4} 
\end{figure}

For the second layer, three qubits are used. The function that will map our data into a quantum state is represented in Fig. \ref{fig:4}. The parameterization used will also be given by $U(\pmb{\theta})$, Eq. \ref{eq:17}. For the measurements we use Eq. \eqref{eq:17.1} with $H_{i}$ defined in Eq. \eqref{eq:17.2}. In this case the last two qubits will be used.

\begin{figure}[H]
    \centering
    \begin{quantikz}[row sep=0.1cm,column sep=0.1cm]
    \lstick{\ket{0}}& \gate[5][1.5cm]{INPUT_{1}}     &  \gate[5][0.5cm]{U_{1}} &  \gate[5][0.5cm]{U_{2}}  & \gate[5][0.5cm]{U_{3}} & \gate[5][0.5cm]{U_{4}}&\\
    \lstick{\ket{0}}&                          &                       &                        &                      &                     & \\
    \lstick{\ket{0}}&                          &                       &                        &                      &                     & \meter{}\\
    \lstick{\ket{0}}&                         &                       &                        &                      &                     & \meter{}\\
    \lstick{\ket{0}}&                          &                       &                        &                      &                     & \meter{} \\
    \end{quantikz}
    $\longrightarrow$\begin{quantikz}[row sep=0.1cm,column sep=0.1cm]
    \lstick{\ket{0}}& \gate[3][1.5cm]{INPUT_{2}}     &  \gate[3][0.5cm]{U_{1}} &  \gate[3][0.5cm]{U_{2}}  & \gate[3][0.5cm]{U_{3}} & \gate[3][0.5cm]{U_{4}}&\\
    \lstick{\ket{0}}&                          &                       &                        &                      &                     & \meter{}\\
    \lstick{\ket{0}}&                          &                       &                        &                      &                     & \meter{}\\
    \end{quantikz}
    \caption{Model 1: Quantum-Classical Hybrid Neural Network.}
    \label{fig:1}
\end{figure}
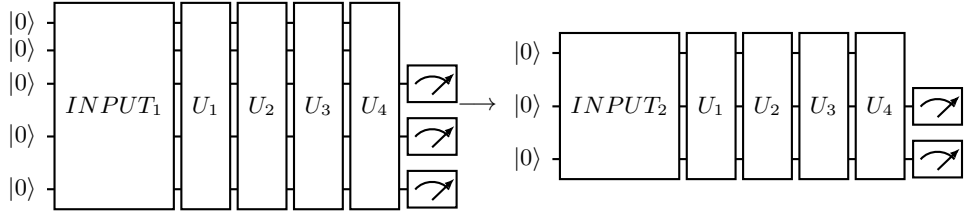

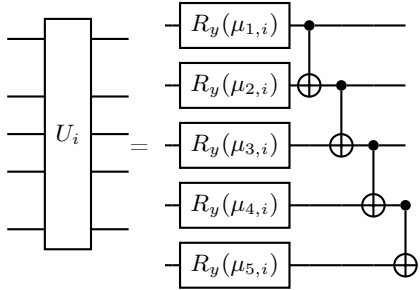
\begin{figure}[H]
    \centering
    \begin{quantikz}
    \qw&\gate[5][0.5cm]{U_{i}}&\qw\\
    \qw&&\qw\\
    \qw&&\qw\\
    \qw&&\qw\\
    \qw&&\qw\\
    \end{quantikz}
    =\begin{quantikz}[row sep=0.2cm,column sep=0.1cm]
    \qw&\qw&    \gate[1]{ R_{y}(\mu_{1,i}) }  & \ctrl{1} & \qw     &  \qw      &  \qw  \\
    \qw&\qw&    \gate[1]{ R_{y}(\mu_{2,i}) }  & \targ{}  & \ctrl{1}&  \qw      &   \qw \\
    \qw&\qw&    \gate[1]{ R_{y}(\mu_{3,i}) }  & \qw      & \targ{} &  \ctrl{1} &    \qw\\
    \qw&\qw&    \gate[1]{ R_{y}(\mu_{4,i}) }  & \qw      & \qw     &    \targ{}&   \ctrl{1}  \\
    \qw&\qw&    \gate[1]{ R_{y}(\mu_{5,i}) }  & \qw      & \qw     &  \qw      &  \targ{}  \\
    \end{quantikz}
    
    \caption{Parameterization of unitary operators $U_{i}$. See Figure \ref{fig:1}.}
    \label{fig:2}
\end{figure}

\subsubsection{Model 2} 
\label{Modelo 2}

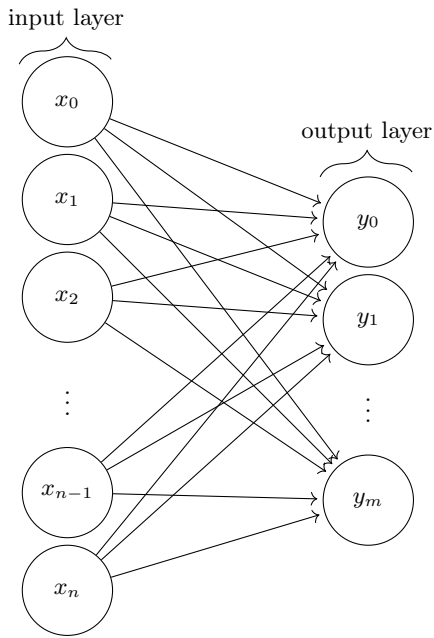
\begin{figure}[H]
    \centering
    \begin{tikzpicture}[shorten >=2pt]
            \tikzstyle{unit}=[draw,shape=circle,minimum size=1.2cm]
            \node[unit](x1) at (0,4){$x_0$};
            \node[unit](x2) at (0,2.7){$x_1$};
            \node[unit](x3) at (0,1.4){$x_2$};
            \node(dots) at (0,0.1){\vdots};
            \node[unit](x4) at (0,-1.2){$x_{n-1}$};
            \node[unit](x5) at (0,-2.5){$x_n$};
           
            \node[unit](x6) at (4,2.4){$y_0$};
            \node[unit](x7) at (4,1.1){$y_1$};
            \node(dots) at (4,0.0){\vdots};
            \node[unit](x8) at (4,-1.3){$y_m$};
           
           \draw[->] (x1) -- (x6);
           \draw[->] (x1) -- (x7);
           \draw[->] (x1) -- (x8);
           
           \draw[->] (x2) -- (x6);
           \draw[->] (x2) -- (x7);
           \draw[->] (x2) -- (x8);
           
           \draw[->] (x3) -- (x6);
           \draw[->] (x3) -- (x7);
           \draw[->] (x3) -- (x8);
           
           \draw[->] (x4) -- (x6);
           \draw[->] (x4) -- (x7);
           \draw[->] (x4) -- (x8);
          
           \draw[->] (x5) -- (x6);
           \draw[->] (x5) -- (x7);
           \draw[->] (x5) -- (x8);
          
           \draw [decorate,decoration={brace,amplitude=10pt},xshift=-4pt,yshift=0pt] (-0.5,4.5) -- (0.75,4.5) node [black,midway,yshift=+0.6cm]{input layer};
           \draw [decorate,decoration={brace,amplitude=10pt},xshift=-4pt,yshift=0pt] (3.5,3) -- (4.75,3) node [black,midway,yshift=+0.6cm]{output layer};
        \end{tikzpicture}
           
    \caption{Linear Layer.}
    \label{fig:11}
\end{figure}

The second model will be built using three layers. The first and third being classical layers, with a quantum
second layer. In Eq. \eqref{eq:100.1} we defined the classical layer that we will use. This consists of an operation that takes an input $\pmb{x}$ of size $n$ into an output $\pmb{y}$ of size $m$, as illustrated in Fig. \ref{fig:11}. In this study, as we will work with data from the dataset MNIST, which consists of images of dimension $28\times28$, the input of the first layer have size $n=784$. The output data is encoded in the quantum layer using the parameterization model presented in Fig. \ref{fig:4}, that is, for each qubit a value is encoded. Then the output of the first layer is equal in size to the number of qubits used in the quantum layer. For nonlinearity, the hyperbolic tangent function will be used.

\begin{figure}[H]
    \centering
     \begin{quantikz}[row sep=0.4cm,column sep=0.4cm]
        \lstick{\ket{0}}& \gate[4][1.5cm]{INPUT_{2}}     &  \gate[4][0.6cm]{U_{1}} &  \gate[4][0.6cm]{U_{2}}  & \gate[4][0.6cm]{U_{3}} & \gate[4][0.6cm]{U_{4}}&\meter{}\\
        \lstick{\ket{0}}&                          &                       &                        &                      &                     & \meter{}\\
        \lstick{\ket{0}}&                          &                       &                        &                      &                     & \meter{}\\
        \lstick{\ket{0}}&                          &                       &                        &                      &                     & \meter{}\\
    \end{quantikz}
    \caption{Quantum Layer.}
    \label{fig:122}
\end{figure}
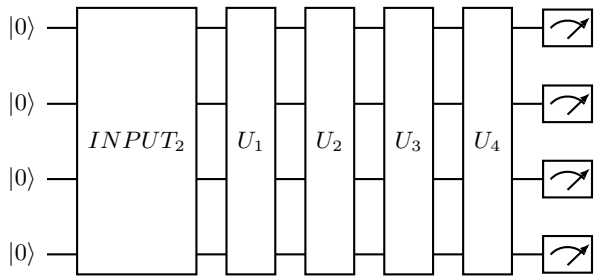

The second layer, which is a quantum layer, is represented in Fig. \ref{fig:122}. In this layer we use four qubits. Its parameterization will be given by $U(\pmb{\theta})$, Eq. \eqref{eq:17}, with each $U_{i}$ represented in Fig. \ref{fig:2}. For measurements, we use Eq. \eqref{eq:17.1} with the Hermitian operator defined in Eq. \eqref{eq:17.2}, with all qubits measured individually. The output of this quantum layer is used as input to a third layer, which is a classical layer, just like the first layer, with the difference being its size. The input of this third layer have dimension equal to the number of qubits used in the quantum layer and output $m=2$.

\subsection{Training} \label{Treinamento}

Given $ \mathcal{D} :=\{ x_{i},\bar{y}_{i} \}_{i=1}^{n} $, a dataset, and defined the HQCNN model that shall be used, the training consists of an iterative method where, given an input $x_{i}$ to the model, it returns an output $y_{i}$. Comparing this output to the desired output $\bar{y}_{i}$, we can compute the performance of our model using the cost function of Eq. \eqref{eq:200}. This process is performed iteratively for all data in the set $\mathcal{D}$, several times, or, as it is commonly called, for several epochs. During this process, we aim to minimize the cost function, that is to say, we want to obtain
\begin{equation}
    \Delta_{opt} = argmin_{\Delta}C(\Delta),
\end{equation}
where $\Delta_{opt}$ is the set of optimal parameters. To obtain $\Delta_{opt}$, at each iteration the $\Delta$ parameters are updated generally using the gradient descent method or its variants. This method updates the parameters using the gradient of the cost function of Eq. \eqref{eq:200}. For this, we must use the chain rule. For example, let us consider the first model. Given an input
$x_{i} = ( x_{i}^{1},x_{i}^{2},...,x_{i}^{n} ) $ and its respective desired output $\bar{y} _{i} = ( \bar{y}_{i}^{1},\bar{y}_{i}^{2},...,\bar{y}_{i}^{N } ) $, the cost function is given by
\begin{equation}
    C(\pmb{\theta}_{1},\pmb{\theta}_{2}) = \frac{1}{N}\sum_{j=1}^{N}( \mathcal{L}_{2}(\mathcal{L}_{1}(x_{i},\pmb{\theta}_{1}), \pmb{\theta}_{2} )^{j} - \bar{y}_{i}^{j} )^{2},
\end{equation}
where $\mathcal{L}_{1}$ is the first layer with input $x_{i}$ and parameters $\pmb{\theta}_{1}$ and $\mathcal{L}_{2}$ is the second layer with input given by the output of the first layer and with parameters $\pmb{\theta}_{2}$. Thus, the gradient of Eq. \eqref{eq:200} in relation to the parameters $\pmb{\theta}_{1}$ and $\pmb{\theta}_{2}$ will be
\begin{align}
    \nabla_{\pmb{\theta}_{2}} C(\pmb{\theta}_{1},\pmb{\theta}_{2}) & = \frac{2}{N}\sum_{j}^{N}( \mathcal{L}_{2}(\mathcal{L}_{1}(x_{i},\pmb{\theta}_{1}), \pmb{\theta}_{2} )^{j} - \bar{y}_{i}^{j} )\nabla_{\pmb{\theta}_{2}}\mathcal{L}_{2}(\mathcal{L}_{1}(x_{i},\pmb{\theta}_{1}), \pmb{\theta}_{2} )^{j}, \label{eq:201} \\
    \nabla_{\pmb{\theta}_{1}}  C(\pmb{\theta}_{1},\pmb{\theta}_{2}) & = \frac{2}{N}\sum_{j}^{N}( \mathcal{L}_{2}(\mathcal{L}_{1}(x_{i},\pmb{\theta}_{1}), \pmb{\theta}_{2} )^{j} - \bar{y}_{i}^{j} )\nabla_{\mathcal{L}_{1}}\mathcal{L}_{2}(\mathcal{L}_{1}(x_{i},\pmb{\theta}_{1}), \pmb{\theta}_{2} )^{j}\nabla_{\pmb{\theta}_{1}}\mathcal{L}_{1}(x_{i},\pmb{\theta}_{1}), \label{eq:202}
\end{align}
where in Eq. \eqref{eq:202} the term $\nabla_{\mathcal{L}_{1}}$ indicates that we must obtain the derivatives with respect to the input of the second layer. Using gradient descent as an example, from Eqs. \eqref{eq:201} and \eqref{eq:202} we have that the new parameters will be
\begin{equation}
    \pmb{\theta}_{1}^{t+1} = \pmb{\theta}_{1}^{t}-\eta\nabla_{\pmb{\theta}_{1}} C(\pmb{\theta}_{1},\pmb{\theta}_{2})
\end{equation}
and
\begin{equation}
    \pmb{\theta}_{2}^{t+1} = \pmb{\theta}_{2}^{t}-\eta\nabla_{\pmb{\theta}_{2}} C(\pmb{\theta}_{1},\pmb{\theta}_{2})
\end{equation}
where $t$ represents the epoch and $\eta$ the learning rate.

Let us consider the term $\nabla_{\pmb{\theta}_{1}}\mathcal{L}_{1}(x_{i},\pmb{\theta}_{1})$ in Eq. \eqref{eq:202}. Once this is a black box function, we can define $J(\pmb{\theta}_{1}) := \mathcal{L}_{1}(x_{i},\pmb{\theta} _{1})$. Here we can see that, up to an index, we can rewrite our quantum layer using Eq. \eqref{eq:1}. So, we can use Eq. \eqref{eq:7} to estimate the gradient.

In Eq. \eqref{eq:202}, we see that we must also obtain the gradient with respect to the input data of the second layer. Again, we can consider the quantum layer to be a black box function with parameters given by $\mathcal{L}_{2}(\mathcal{L}_{1}(x_{i},\pmb{\theta}_{1}), \pmb{ \theta}_{2})$. Therefore, we can again use Eq. \eqref{eq:1} to describe this layer. Then the gradient can be estimated using Eq. \eqref{eq:7}.

\subsection{Barren Plateaus} \label{Barren_Plateaus}

The optimization of quantum circuits is done,  in general, using the gradient in relation to its parameters. However, a current problem with this procedure is the phenomenon known as gradient vanishing, or barren plateaus. Given a function
\begin{equation}
C = Tr[ O U(\pmb{\theta})|\pmb{x} \rangle  \langle  \pmb{x}|U(\pmb{\theta})^{\dagger} ],
\end{equation}
if $U(\pmb{\theta}) $ is a $2$-\textit{design}, then we have that
\begin{equation}
\langle \partial_{k}C \rangle  = 0 \text{ and } Var[\langle \partial_{k}C \rangle] \approx 2^{-n},
\end{equation}
where $n$ is the number of qubits. From the Chebyshev inequality,
\begin{equation}
    Pr( |\partial_{k}C| \geqslant \delta )  \leqslant \frac{ Var[\langle \partial_{k}C \rangle]  }{\delta^{2}}, \label{eq:3003}
\end{equation}
we have that the probability that $\partial_{k}C$ deviates from its mean, $ \langle \partial_{k}C \rangle = 0 $ , by a value $\delta$ will tend to zero as the number of qubits increase.

Furthermore, results from the literature show that other factors also influence this phenomenon such as the choice of cost function \cite{barrenPlateaus_1}, expressiveness of parameterization \cite{barrenPlateaus_2}, noise \cite{barrenPlateaus_3} and entanglement \cite{barrenPlateaus_4,barrenPlateaus_5}. It is also shown that gradient vanishing is present in gradient free optimization methods \cite{barrenPlateaus_6}. Currently, a large number of methods to mitigate this problem are being proposed \cite{FRIEDRICH,BR_initialization_strategy,BR_Large_gradients_via_correlation,BR_LSTM,BR_layer_by_layer}.

\begin{figure}[H]
\centering
    \includegraphics[scale=0.7]{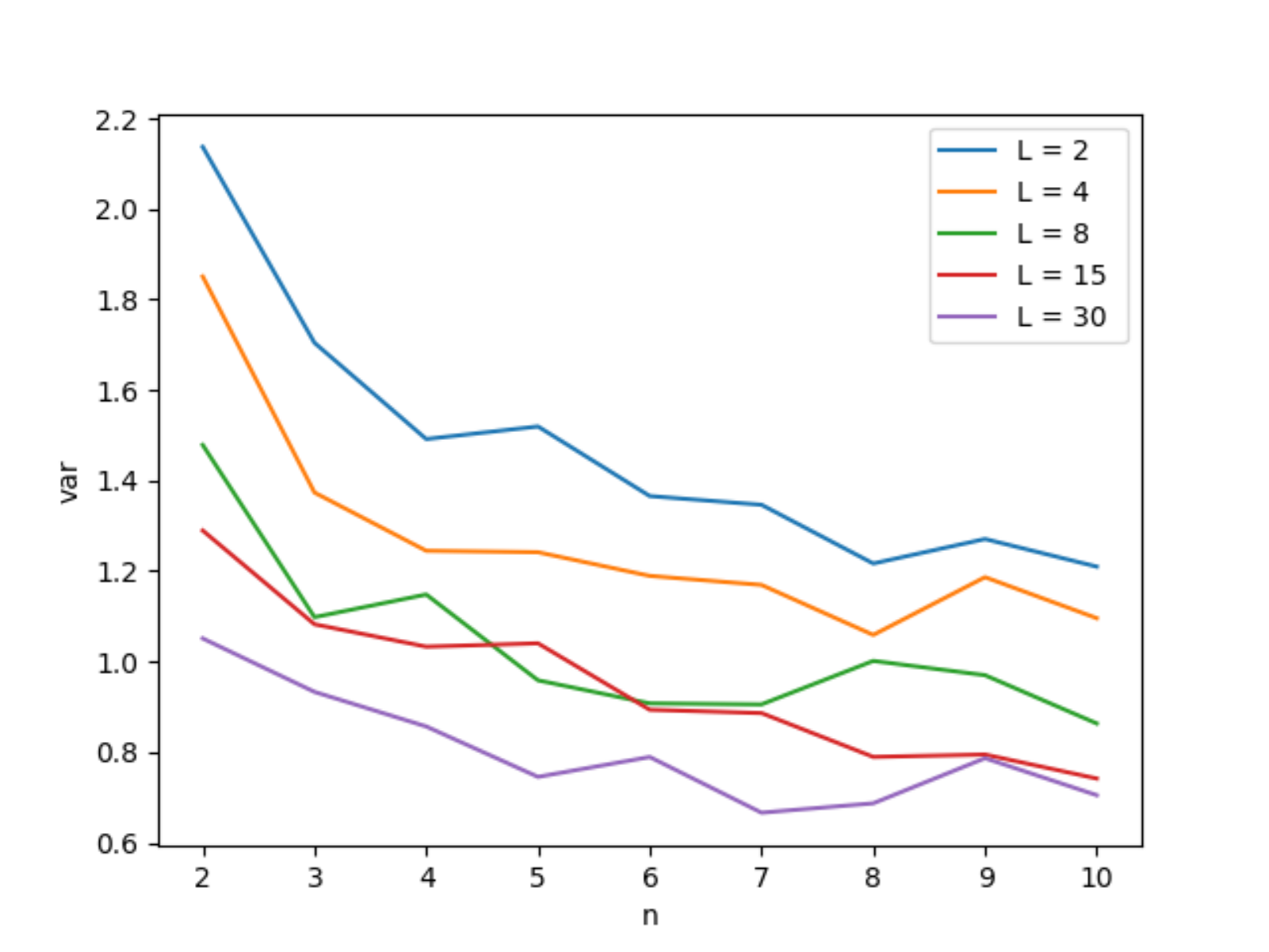}
    \caption{In this figure, $n$ is the number of qubits used. For the input data, $ \pmb{x} = ( \pi/4,\pi/4,\pi/4,...,\pi/4 ) $ was used. We can see that as the number of qubits increases, the variance tends to decrease. Furthermore, we can also see that the variance decreases as the number of layers of the parameterization increases. Thus, we see that the phenomenon of gradient disappearance is also present in the evolution strategy method.}
    \label{fig:variance}
\end{figure}

In Fig. \ref{fig:variance}, we showed experimentally that this phenomenon is also observed when using the ES method. To obtain these results, we use the cost function
\begin{equation}
    C = \frac{1}{n}\sum_{i=1}^{n}Tr[H_{i} U(\pmb{\theta})\rho U(\pmb{\theta})^{\dagger}  ],
\end{equation}
with $U(\pmb{\theta}) $ defined in Eq. \eqref{eq:17} and $H_{i}$ is defined in Eq. \eqref{eq:17.2}. To encode the input data, we use the parameterization shown in Fig. \ref{fig:4}.

\section{Experiments/Simulations} 
\label{Experiments}

For this work, we will use Qiskit \cite{19} and Pytorch \cite{20} to build our models, with the Qiskit package integrated into Pytorch. With this, some steps such as the application of the chain rule to obtain the derivatives in Eqs. \eqref{eq:201} and \eqref{eq:202} and parameters optimization are done automatically by Pytorch. For training, $2000$ training images are used, half referring to zero digit images and half to the one digit images. For validation, $200$ images are used, again half of each type. We also vary the value of the learning rate to see how the cost function behaves. Also, we perform the same experiment $N$ times for obtaining the statistics. So, we can see how the ES method behaves for different initializations. Also, for this study we used $\sigma = \frac{\pi}{24}$ for all cases. As $\sigma$ is a hyperparameter that must be defined when starting the training, as well as the number of epochs, learning rate and $\lambda$, your choice is free, so we use only one value for $\sigma$ , since our objective is to analyze whether this method is capable of performing the optimization of the parameters and not which is the best set of hyperparameters. However, we must emphasize that there is no restriction on this choice. For the first experiments, we used 
\begin{equation}
\lambda = 4+3\log(p),
\end{equation}
where $p$ is the number of parameters of the respective quantum layer. For the layers where we should get the gradient for the input data, we use the highest value of $\lambda$. That is, given the value of $\lambda$ for the layer's input data and parameters, we will use the largest of them.

Search gradients give us freedom to select the number of function evaluations. The second experiment explores this by using the same values for $\eta$ used in experiment one, but now with differing evaluation numbers, $\lambda=2$, $\lambda=4$, and $\lambda=6$. Both experiments use HQCNN models applied to part of the MNIST dataset.

\section{Results} \label{Results}

The first experiment fed a subset of the MNIST dataset into models 1 and 2. Optimization was performed using Adam with a variable learning rate, $\eta$.  For all experiments, 100 epochs were used for training. For each learning rate value, the same experiment was repeated four times. At each new training, the parameters were randomly initialized. With this we can see how the ES method behaves at each startup.

\begin{figure}[H]
  \centering
        \includegraphics[scale=0.40]{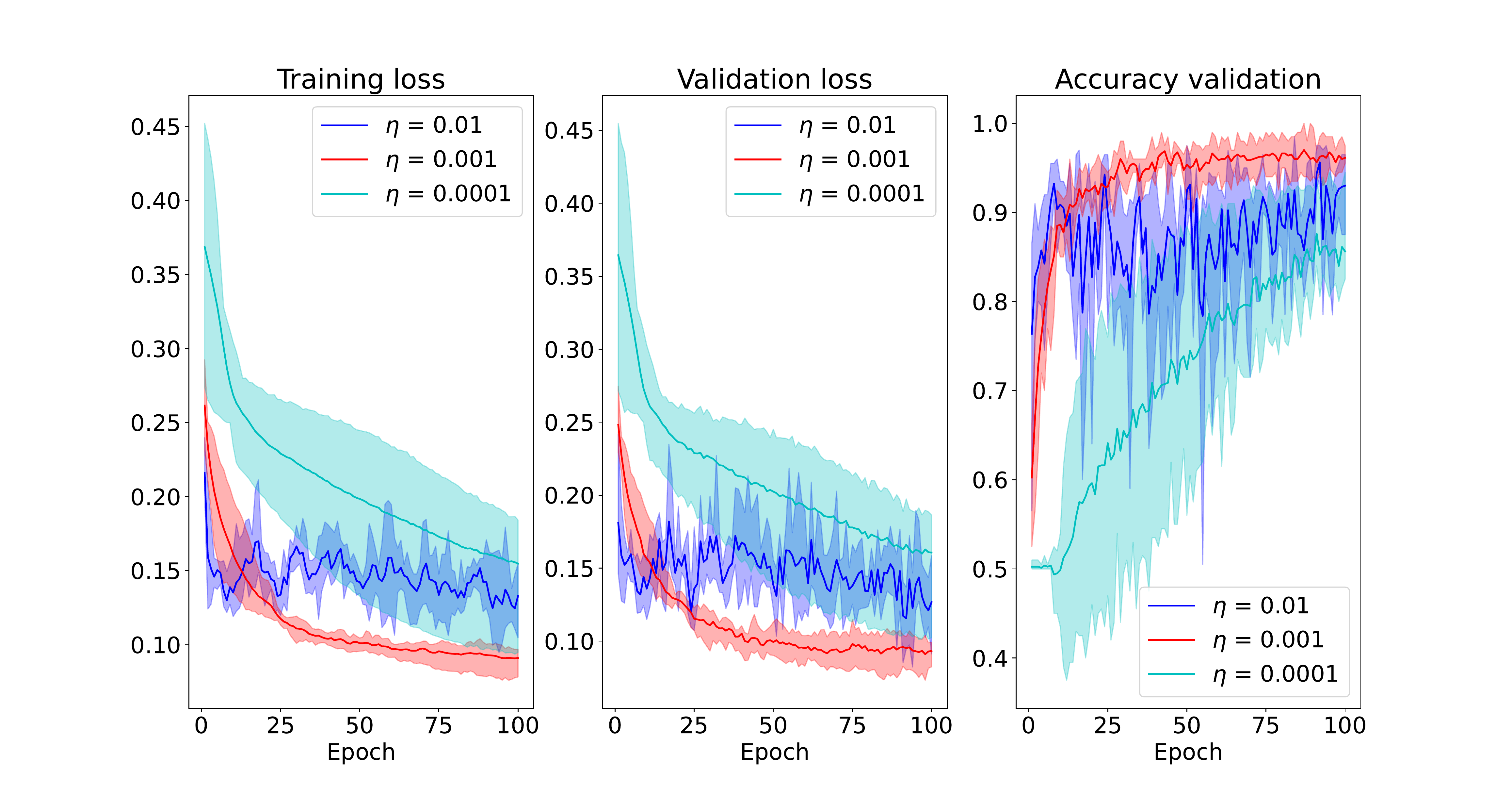}
        \caption{ Result for Model 1. In darker colors the means for $N=4$ experiments are shown. The lighter colors represent values between the minimum and maximum values. To obtain the gradient estimate, the  Algorithm \ref{alg:1} was used. In order to be able to update the parameters of the first layer, we must get the gradient from the input data of the second layer. For this we can also use the  Algorithm \ref{alg:1}, with $\pmb{\theta}$ being the input data in this case.}
       \label{fig:5}
\end{figure}

In Fig. \ref{fig:5}, we can see that for $\eta = 0.01$  the neural network becomes stuck in a local minima. For $\eta=0.001$, we see that the neural network can learn as the epochs go by. For $\eta=0.0001$, we see that for this number of epochs used, the neural network performed worse than in the other cases. But we can see that even as the epochs passed, the network was able to learn without getting stuck in a local minimum, as was the case for $\eta=0.01$.

\begin{figure}[H]
\centering
        \includegraphics[scale=0.40]{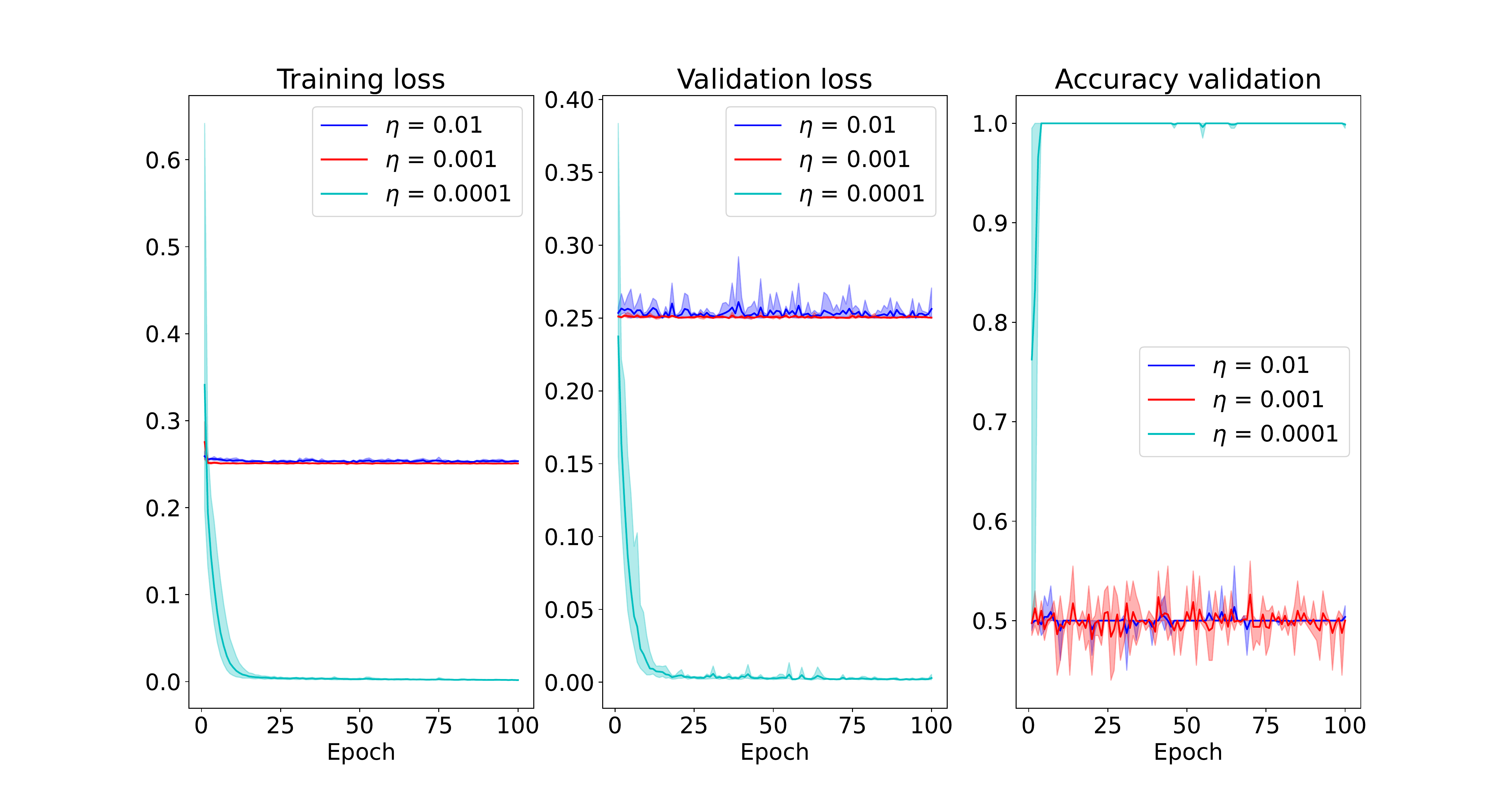}
        \caption{ Results for Model 2. The means for $N=4$ experiments are shown in darker colors, while the lighter colors represent values between the minimum and maximum values. Algorithm \ref{alg:1} was used to obtain the gradient estimate. To update the parameters of the first layer, we must get the gradient from the input data of the second layer. For this we can also use the Algorithm \ref{alg:1}, with $\pmb{\theta}$ being the input data in this case. }
        \label{fig_6}
\end{figure}

Results shown in Fig. \ref{fig_6} use Model 2 whereas results in Fig. \ref{fig:5} use Model 1. Both models used the same training and validation sets. Again we use the Adam optimizer with variable learning rate. Unlike the previous model, we see that for both $\eta=0.01$ and $\eta=0.001$, the neural network was stuck in a local minimum, being unable to learn. The neural network was only able to learn when we used $\eta=0.0001$, where we can see that it quickly converged to the desired minimum.

\begin{figure}[H]
        \centering
        \includegraphics[scale=0.42]{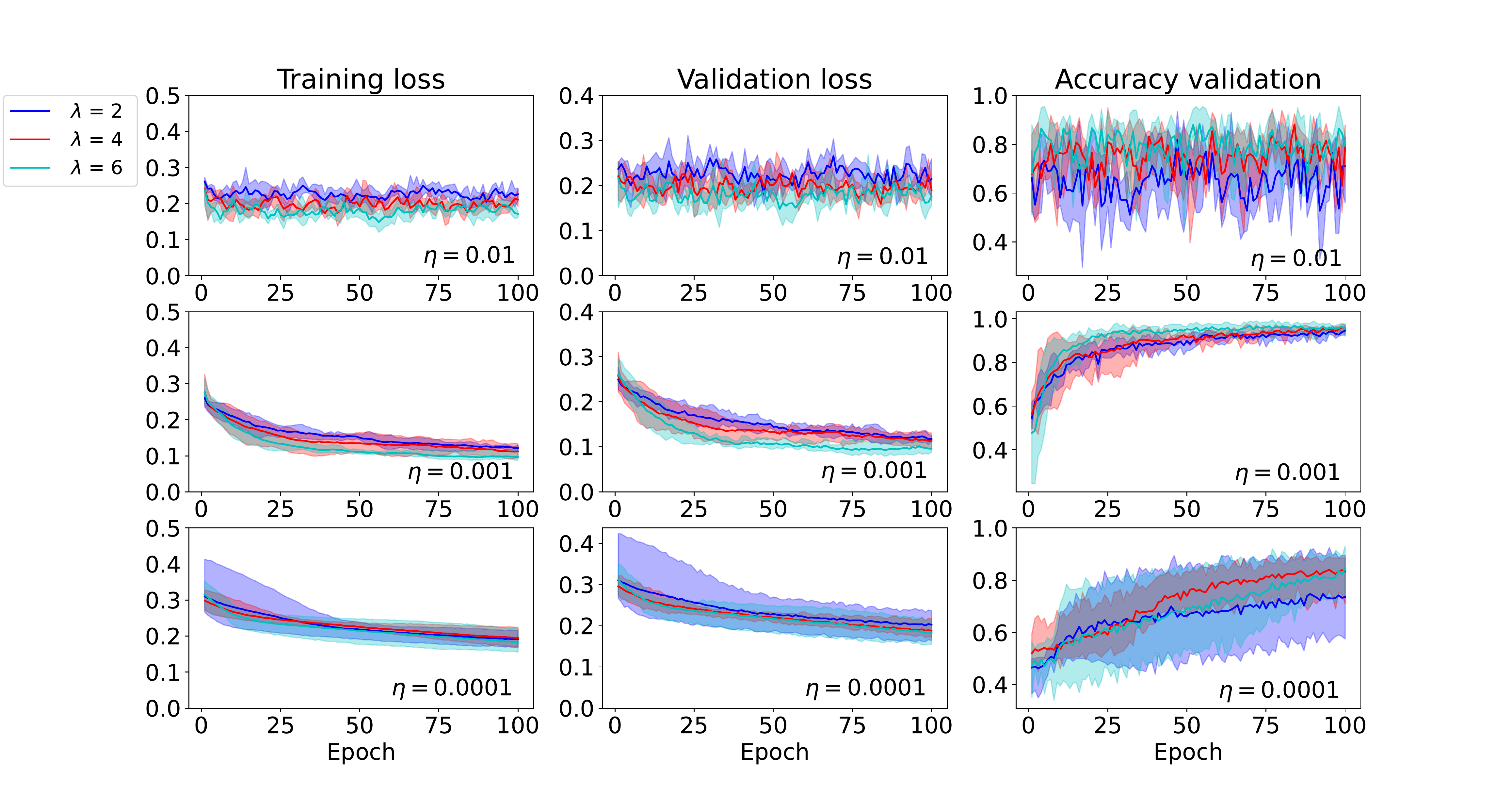}
        \caption{Results for Model 1 with variable $\lambda$ and learning rate $\eta$. In darker colors, the averages for $N=4$ experiments are shown. The lighter colors represent values between the minimum and maximum values. }
        \label{fig:5.1}
\end{figure}

\begin{figure}[H]
\centering
        \includegraphics[scale=0.42]{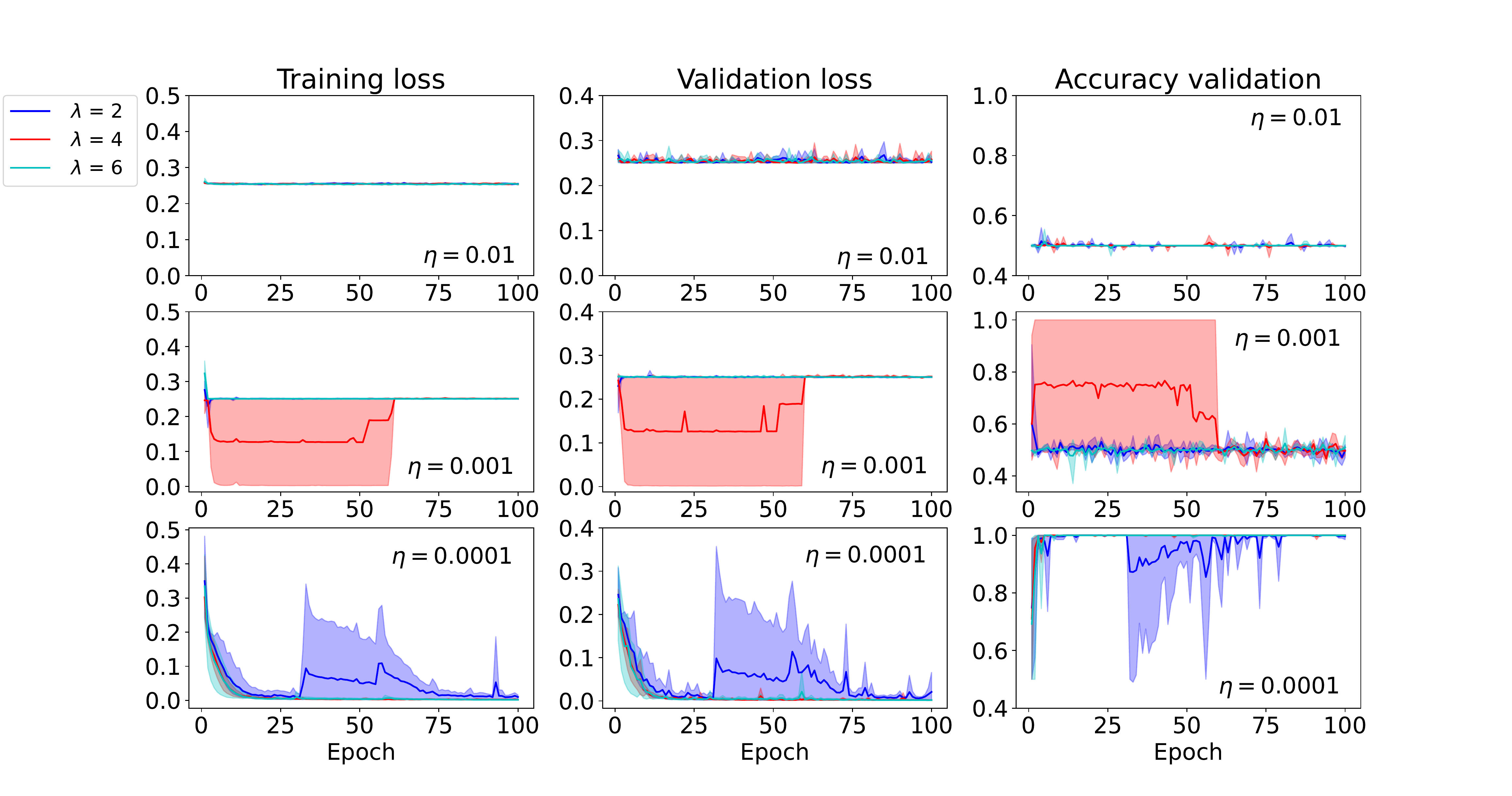}
        \caption{ Results for Model 2 with variable $\lambda$ and learning rate $\eta$. In darker colors the averages for $N=4$ experiments are shown. The lighter colors represent values between the minimum and maximum values.}
        \label{fig:5.2}
\end{figure}

In the next graphs, in Figs. \ref{fig:5.1} and \ref{fig:5.2}, Models 1 and 2 were used, respectively, with the data set defined as in the previous experiments, and with a variable learning rate. In these two cases, the difference lies in the number of times the cost function is evaluated, that is to say, the value of $\lambda$ that is used to estimate the gradient, Eq. \eqref{eq:7}. In the case of two quantum layers, Fig. \ref{fig:5.1}, we see that using $\eta=0.01$ the neural network was not able to learn for any value of $\lambda$. For $\eta=0.001$, its behavior was similar for all values of $\lambda$. As for $\eta=0.0001$, its behavior was not better than using $\eta=0.001$, but we can see that the neural network was able to learn over the epochs for all values of $\lambda$.

For the second model we se that performance was only satisfactory for $\eta = 0.0001$. For $\eta = 0.01$, and $\eta = 0.001$, networks became trapped in local minima for all values of $\lambda$ that we explored. With this, we can see experimentally that the performance of the ES method applied in hybrid quantum-classical models, where both classical and quantum layers are used, Model 2 has greater dependence on its hyperparameters.

\section{Conclusions} \label{Conclusions}
In this article we introduced the use of the method called evolution strategy in the optimization of classical-quantum hybrid neural networks. We showed that this method is strongly influenced by the hyperparameters $\eta$ and $\lambda$. However, this is a promissing method for optimizing hybrid models, once the appropriate hyperparameters are provided. Therefore, for future work it would be interesting to study methods for selecting hyperparameters e.g. through standard hyperparameter optimization techniques such as Bayesian optimization. In addition, another possible topic for research is the use of natural gradients, in what is known as a natural evolution strategy. In this case it would be interesting to evaluate if the performance of the model will be better in relation to the method  studied in this work. Finally, as we observed, although the method applied in this article does not directly use the derivatives in relation to the parameters of the quantum circuit to optimize them, it still suffers from the problem of vanishing gradients. So, in the future one can investigate the use of methods such as the one introduced in Ref. \cite{BR_layer_by_layer}, where the optimization of the parameters is done layer by layer to mitigate this problem, and also to see if by performing the training layer by layer the performance will be better than the performance obtained in this article.

\begin{acknowledgments}
This work was supported  by the Foundation for Research Support of the State of Rio Grande do Sul (FAPERGS), by the National Institute for the Science and Technology of Quantum Information (INCT-IQ), process 465469/2014-0, and by the National Council for Scientific and Technological Development (CNPq), process 309862/2021-3.
\end{acknowledgments}

\vspace{0.3cm}

\textbf{Data availability} The Qiskit/Pytorch code used for implementing the simulations to obtain the data used in this article is available upon request to the authors.

\vspace{0.3cm}

\textbf{Declaration of competing interest}  The authors declare that they have no known competing financial interests or personal relationships that could have appeared to influence the work reported in this paper.



\begin{thebibliography}{4}


\bibitem{computer_vision_1} K. He, X. Zhang, S. Ren, and J. Sun, Deep residual learning for image recognition, Proceedings of the IEEE Conference on Computer Vision and Pattern Recognition (2016).

\bibitem{computer_vision_2} C. Szegedy et al., Going deeper with convolutions, Proceedings of the IEEE Conference on Computer Vision and Pattern Recognition (2015).

\bibitem{computer_vision_3} A. Voulodimos, N. Doulamis, A. Doulamis, and E. Protopapadakis, Deep Learning for Computer Vision: A Brief Review, Computational Intelligence and Neuroscience, 2018, e7068349 (2018).

\bibitem{nlp_1} J. Devlin, M.-W. Chang, K. Lee, and  K. Toutanova, Bert: Pre-training of deep bidirectional transformers for language understanding, 
arXiv.1810.04805 (2018).

\bibitem{nlp_2} I. Sutskever, O. Vinyals, and Q. V. Le, Sequence to sequence learning with neural networks, Advances in Neural Information Processing Systems 27 (2014).

\bibitem{drog_discovery} J. Vamathevan et al., Applications of machine learning in drug discovery and development, Nature Reviews Drug Discovery 18, 463 (2019).

\bibitem{dl_astronomia} Rodrigo Carrasco-Davis et al, Deep learning for image sequence classification of astronomical events, Publications of the Astronomical Society of the Pacific, 131, 108006 (2019).

\bibitem{Deep_Chemistry} T.F.G.G. Cova and A.A.C.C. Pais, Deep learning for deep chemistry: optimizing the prediction of chemical patterns, Frontiers in chemistry 7, 809 (2019).

\bibitem{7} S. Garg and G. Ramakrishnan, Advances in quantum deep learning: An overview, 
arXiv.2005.04316 (2020).
\bibitem{7.2} N. A. Nghiem, S. Y.-C. Chen, and T.-C. Wei, A Unified Framework for Quantum Supervised Learning, Phys. Rev. Research 3, 033056 (2021) 2020.
\bibitem{7.3} E. Farhi and H. Neven, Classification with Quantum Neural Networks on Near Term Processors, 
arXiv.1802.06002 (2018).

\bibitem{7.4} F. Tacchino, P. Barkoutsos, C. Macchiavello, I. Tavernelli, D. Gerace, and D. Bajoni, Quantum implementation of an artificial feed-forward neural network,  Quantum Sci. Technol. 5, 044010 (2020). 


\bibitem{7.5} G. Verdon, J. Pye, and M. Broughton, A Universal Training Algorithm for Quantum Deep Learning,  
arXiv.1806.09729 (2018).

\bibitem{7.6}
K. Beer, D. Bondarenko, T. Farrelly, T. J. Osborne, R. Salzmann, D. Scheiermann, and R. Wolf, Training deep quantum neural networks, Nature Comm. 11, 808 (2020).

\bibitem{7.7} 
S. Wei, Y. Chen, Z. Zhou, and G. Long, A Quantum Convolutional Neural Network on NISQ Devices, 
arXiv.2104.06918 (2021).

\bibitem{7.8} S. Lloyd, M. Schuld, A. Ijaz, J. Izaac, and N. Killoran, Quantum embeddings for machine learning, 
arXiv.2001.03622 (2020).


\bibitem{quantum_model_multilayer_perception} C. Shao, A quantum model for multilayer perceptron, 
arXiv.1808.10561 (2018).

\bibitem{qunatum_Convolutional} S.J. Wei, Y.H. Chen, Z.R. Zhou, and G.L. Long, A quantum convolutional neural network on NISQ devices, AAPPS Bull. 32, 2 (2022).

\bibitem{kernel_methods} M. Schuld, Supervised quantum machine learning models are kernel methods, 
arXiv.2101.11020 (2021).

\bibitem{hybrid_1} J. Liu et al., Hybrid quantum-classical convolutional neural networks, Sci. China Phys. Mech. Astron. 64, 290311 (2021).

\bibitem{hybrid_2} Y. Liang, W. Peng, Z.-J. Zheng, O. Silv\'en, and G. Zhao, A hybrid quantum–classical neural network with deep residual learning, Neural Networks 143, 133 (2021).

\bibitem{hybrid_3} R. Xia and S. Kais, Hybrid quantum-classical neural network for calculating ground state energies of molecules, Entropy 22, 828 (2020).

\bibitem{hybrid_4} E. H. Houssein, Z. Abohashima, M. Elhoseny, and W. M. Mohamed, Hybrid quantum convolutional neural networks model for COVID-19 prediction using chest X-Ray images, Journal of Computational Design and Engineering 9, 343 (2022).

\bibitem{maria} M. Schuld, V. Bergholm, C. Gogolin, J. Izaac, and N. Killoran, Evaluating analytic gradients on quantum hardware, Phys. Rev. A 99, 032331 (2019).




\bibitem{es_1}Rechenberg, Ingo. "Evolutionsstrategien." Simulationsmethoden in der Medizin und Biologie. Springer, Berlin, Heidelberg, 1978. 83-114.

\bibitem{es_2} Schwefel, H. P. (1977). Numerische Optimierung von Computer-Modellen mittels der Evolutionsstrategie: mit einer vergleichenden Einführung in die Hill-Climbing-und Zufallsstrategie (Vol. 1). Basel: Birkhäuser.

\bibitem{10} D. Wierstra et al., Natural evolution strategies, Journal of Machine Learning Research 15, 949 (2014).

\bibitem{11} G. E. Crooks,  Gradients of parameterized quantum gates using the parameter-shift rule and gate decomposition, 
arXiv.1905.13311 (2019). 

\bibitem{12.3} T. Salimans et al., Evolution strategies as a scalable alternative to reinforcement learning. arXiv:1703.03864 (2017).

\bibitem{12.0} J. Yao, M. Bukov, and L. Lin, Policy Gradient based Quantum Approximate Optimization Algorithm, 
arXiv.2002.01068 (2020).   

\bibitem{12.1} A. Anand, M. Degroote, and A. Aspuru-Guzik, Natural evolutionary strategies for variational quantum
computation, Mach. Learn.: Sci. Technol. 2, 045012 (2021).

\bibitem{12.2} M. Wilson, S. Stromswold, F. Wudarski, S. Hadfield, N. M. Tubman, and E. Rieffel, Optimizing quantum heuristics with meta-learning, Quantum Mach. Intell. 3, 13 (2021).

\bibitem{dataEncoder_1} M. Schuld, R. Sweke, and J. J. Meyer, The effect of data encoding on the expressive power of variational quantum machine
learning models, Phys. Rev. A 103, 032430 (2021).

\bibitem{dataEncoder_2} A. P\'erez-Salinas, A. Cervera-Lierta, E. Gil-Fuster, and J. I. Latorre, Data re-uploading for a universal quantum classifier, Quantum 4, 226 (2020).




\bibitem{dataEncoder_4} R. LaRose and B. Coyle, Robust data encodings for quantum classifiers, Phys. Rev. A 102, 032420 (2020).

\bibitem{barrenPlateaus_1} M. Cerezo, A. Sone, T. Volkoff, L. Cincio, and P. J. Coles, Cost function dependent barren plateaus in shallow parametrized quantum circuits, Nature Communications 12, 1 (2021).

\bibitem{barrenPlateaus_2} Z. Holmes, K. Sharma, M. Cerezo, and P. J. Coles, Connecting ansatz expressibility to gradient magnitudes and barren plateaus, PRX Quantum 3, 010313 (2022).

\bibitem{barrenPlateaus_3} S. Wang et al., Noise-induced barren plateaus in variational quantum algorithms, Nature Communications 12, 6961 (2021).

\bibitem{barrenPlateaus_4} C. O. Marrero, M. Kieferov\'a, and N. Wiebe, Entanglement-induced barren plateaus, PRX Quantum 2, 040316 (2021).

\bibitem{barrenPlateaus_5} T. L. Patti, K. Najafi, X. Gao, and S. F. Yelin, Entanglement devised barren plateau mitigation, Phys. Rev. Research 3, 033090 (2021).

\bibitem{barrenPlateaus_6} A. Arrasmith, M. Cerezo, P. Czarnik, L. Cincio, and P. J. Coles, Effect of barren plateaus on gradient-free optimization, Quantum 5, 558 (2021).

\bibitem{FRIEDRICH} L. Friedrich and J. Maziero, Avoiding Barren Plateaus with Classical Deep Neural Networks, Phys. Rev. A 106, 042433 (2022).

\bibitem{BR_initialization_strategy} E. Grant, L. Wossnig, M. Ostaszewski, and M. Benedetti, An initialization strategy for addressing barren plateaus in parametrized quantum circuits, Quantum 3, 214 (2019).

\bibitem{BR_Large_gradients_via_correlation} T. Volkoff and P. J. Coles, Large gradients via correlation in random parameterized quantum circuits, Quantum Science and Technology 6, 025008 (2021).

\bibitem{BR_LSTM} G. Verdon et al., Learning to learn with quantum neural networks via classical neural networks, arXiv.1907.05415 (2019).

\bibitem{BR_layer_by_layer} A. Skolik, J. R. McClean, M. Mohseni, P. van der Smagt, and M. Leib, Layerwise learning for quantum neural networks, Quantum Machine Intelligence 3, 5 (2021).

\bibitem{19} D. C. McKay et al., Qiskit Backend Specifications for OpenQASM and OpenPulse Experiments, 
arXiv.1809.03452 (2018).

\bibitem{20} A. Paszke et al., PyTorch: An Imperative Style, High-Performance Deep Learning Library, 
arXiv.1912.01703 (2019).

\end{thebibliography}
\end{document}